\documentclass[conference]{IEEEtran}
\IEEEoverridecommandlockouts
\usepackage{cite}
\usepackage{amsmath,amssymb,amsfonts}
\usepackage{algorithmic}
\usepackage{algorithm}
\usepackage{graphicx}
\usepackage{textcomp}
\usepackage{xcolor}
\def\BibTeX{{\rm B\kern-.05em{\sc i\kern-.025em b}\kern-.08em
		T\kern-.1667em\lower.7ex\hbox{E}\kern-.125emX}}

\usepackage{booktabs}

\usepackage{tabularx}
\usepackage{array}
\usepackage{graphicx}
\usepackage{float}
\usepackage{subfig}

\makeatletter
\newcommand{\linebreakand}{%
\end{@IEEEauthorhalign}
\hfill\mbox{}\par
\mbox{}\hfill\begin{@IEEEauthorhalign}
}
\makeatother

\begin{document}
\title{Controlled Evolution-Based Day-Ahead Robust Dispatch Considering Frequency Security with Frequency Regulation Loads and Curtailable Loads\\
\thanks{This work is supported by Science and Technology Foundation of State Grid Corporation of China (Grant Number: 5100-202355765A-3-5-YS).}
}
\author{\IEEEauthorblockN{Kai Kang, Xiaoyu Peng}
\IEEEauthorblockA{\textit{Department of Electrical Engineering,}\\ 
\textit{Tsinghua University}\\
	Beijing, China \\
kk21@tsinghua.org.cn }
\and
\IEEEauthorblockN{Kui Luo}
\IEEEauthorblockA{\textit{China Electric Power Research Institute} \\
	Beijing, China \\
	luokui21@163.com
	}
\and
\IEEEauthorblockN{Xi Ru, Feng Liu*}
\IEEEauthorblockA{\textit{Department of Electrical Engineering,}\\ 
	\textit{Tsinghua University} \\
Beijing, China \\
lfeng@tsinghua.edu.cn}
}
\maketitle
\begin{abstract}
With the extensive integration of volatile and uncertain renewable energy, power systems face significant challenges in primary frequency regulation due to instantaneous power fluctuations. However, the maximum frequency deviation constraint is inherently non-convex, and commonly used two-stage dispatch methods overlook causality, potentially resulting in infeasible day-ahead decisions. This paper presents a controlled evolution-based day-ahead robust dispatch method to address these issues. First, we suggest the convex relaxation technique to transform the maximum frequency deviation constraint to facilitate optimization. Then, an evolution-based robust dispatch framework is introduced to align day-ahead decisions with intraday strategies, ensuring both frequency security and power supply reliability. Additionally, a novel controlled evolution-based algorithm is developed to solve this framework efficiently. Case studies on a modified IEEE 14-bus system demonstrate the superiority of the proposed method in enhancing frequency security and system reliability.
\end{abstract}


\section{Introduction}
As resource scarcity pressures mount, the rapid development of renewable energy has become widely embraced in the power industry. However, the stochastic nature of renewable energy output presents significant dispatch challenges: on one hand, the instantaneous fluctuations of renewable power undermine frequency stability \cite{li2021frequency}; on the other hand, the sequential observation of renewable generation demands a dispatch process that respects causality \cite{lv2022multi, kang2024enforcing}.

In primary frequency regulation of power systems, the maximum frequency deviation is a critical indicator of frequency security \cite{gu2020zonal}. Due to its non-convex nature, satisfying this indicator poses substantial challenges for power system decision-making.  To address the frequency issues, literature \cite{restrepo2005unit} integrates frequency security constraints into economic dispatch, focusing solely on quasi-steady-state frequency deviations, and not considering the maximum frequency deviation. Later research \cite{chavez2014governor} proposes constraints for maximum frequency deviation but neglects essential parameters like system damping. To capture non-convex maximum frequency deviations, other works \cite{badesa2019simultaneous} and \cite{zhang2020modeling} utilize piecewise linear approximation with numerous binary variables, substantially increasing the solution complexity significantly.

Currently, power systems generally employ two-stage dispatch methods \cite{qiu2022application}, which assume renewable generation is observable across all intraday periods. However, these assumptions do not satisfy the causal nature of real-world systems. Therefore, two-stage decisions often yield infeasible scheduling outcomes \cite{kang2023day}. In response, multi-stage dispatch methods based on affine policies have been developed \cite{lorca2016multistage,wang2022wasserstein,zhang2020affinely}, but these approaches reduce the feasible set of intraday decision variables significantly, limiting their practical applicability.

This paper investigates the day-ahead robust dispatch problem in power systems. The non-convex maximum frequency deviation indicator is transformed into a set of hyperplane constraints using convex relaxation techniques, enabling efficient optimization without integer variables. Additionally, we introduce a novel evolution-based robust dispatch framework that aligns day-ahead decisions with multi-stage intraday strategies, thereby ensuring causality. To solve this framework, we develop a controlled evolution-based solution algorithm. Testing on a modified IEEE 14-bus system with renewable integration demonstrates that, our method has significant advantages in enhancing frequency security and power supply reliability.

\section{Power system operation model}
\subsection{Outline}
The power system consists of the synchronous generator, renewable generation, energy storage, frequency regulation load,  curtailable load, and load demand. 
Each type of device is represented by a set, denoted as $\Omega_\bullet$, and the number of devices in each set as $\left|\Omega_\bullet\right|$, where $\bullet$ corresponds to SG, RG, ES, FQR, CL, or LD, respectively. The frequency regulation load provides frequency support lasting for seconds \cite{hong2019fast}, while the curtailable load provides hourly time scale power curtailment \cite{liu2020long}. 
The set of periods is denoted as $\Omega_\mathrm{T}$, with $\left|\Omega_\mathrm{T}\right|$ representing the number of periods.

\subsection{Operation Constraints}
Synchronous generator $g\in\Omega_\mathrm{SG}$ in period $t \in \Omega_\mathrm T$ satisfies:
\begin{equation}
	\setlength\abovedisplayskip{0pt}
	\setlength\belowdisplayskip{1pt}
	\label{eq:SG_1}
		(X_{g,t}^\mathrm{ON} - T_g^\mathrm{ON})({x_{g,t}} - {x_{g,t + 1}}) \ge 0
\end{equation}
\begin{equation}
	\label{eq:SG_2}
		(X_{g,t}^\mathrm{OFF} - T_g^\mathrm{OFF})({x_{g,t}} - {x_{g,t + 1}}) \ge 0
\end{equation}
\begin{equation}
	\setlength\abovedisplayskip{0pt}
	\setlength\belowdisplayskip{0pt}
	\label{eq:SG_3}
	{y_{g,t + 1}} - {z_{g,t + 1}} = {x_{g,t + 1}} - {x_{g,t}}
\end{equation}
\begin{equation}
	\setlength\abovedisplayskip{-1pt}
	\setlength\belowdisplayskip{-1pt}
	\label{eq:SG_4}
	p_g^{\min }{x_{g,t}} \le {p_{g,t}} \le p_g^{\max }{x_{g,t}}
\end{equation}
\begin{equation}
	\setlength\abovedisplayskip{0pt}
	\setlength\belowdisplayskip{0pt}
	\label{eq:SG_5}
	{p_{g,t}} - {p_{g,t + 1}} \le {RD_g}{x_{g,t + 1}} + p_g^{\max }(1 - {x_{g,t + 1}})
\end{equation}
\begin{equation}
	\setlength\abovedisplayskip{0pt}
	\setlength\belowdisplayskip{0pt}
	\label{eq:SG_6}
		{p_{g,t + 1}} - {p_{g,t}} \le {RU_g}{x_{g,t}} + p_g^{\max }(1 - {x_{g,t}})
\end{equation} 
where $T_g^\mathrm{ON}$ and $T_g^\mathrm{OFF}$ are the minimum on/off time, $\left[p_g^{\min },p_g^{\max }\right]$ are the power output range, and ${RD_g}$ and $R{U_g}$ are the maximum up/down ramp power. Binary variables ${x_{g,t}}$, ${y_{g,t}}$, and ${z_{g,t}}$ indicate startup, shutdown, and operation states. Variables $X_{g,t}^\mathrm{ON}$, $X_{g,t}^\mathrm{OFF}$, and ${p_{g,t}}$ represent the  operation/non-operation durations and the generator's power output.

Renewable generation $r\in\Omega_\mathrm{RG}$ in period $t$ satisfies:
\begin{equation}
	\setlength\abovedisplayskip{0pt}
	\setlength\belowdisplayskip{0pt}
	\label{eq:RG_1}
	  p_{r,t}^e-p_{r,t}^h \le {p_{r,t}} \le p_{r,t}^e+p_{r,t}^h
\end{equation}
where $p_{r,t}^e$ and $p_{r,t}^h$ denote the expected power output and the deviation of the renewable power ${p_{r,t}}$.

Energy storage $e\in\Omega_\mathrm{ES}$ in period $t$ satisfies:
\begin{equation}
	\setlength\abovedisplayskip{0pt}
	\setlength\belowdisplayskip{0pt}
	\label{eq:ES_1}
	0 \le p_{e,t}^\mathrm{ch} \le p_e^{\max}
\end{equation}
\begin{equation}
	\setlength\abovedisplayskip{0pt}
	\setlength\belowdisplayskip{0pt}
	\label{eq:ES_2}
	0 \le p_{e,t}^\mathrm{dc} \le p_e^{\max}
\end{equation}
\begin{equation}
	\setlength\abovedisplayskip{0pt}
	\setlength\belowdisplayskip{0pt}
	\label{eq:ES_3}
	E_e^{\min} \le {E_{e,t}} \le  E_e^{\max}
\end{equation}
\begin{equation}
	\setlength\abovedisplayskip{0pt}
	\setlength\belowdisplayskip{0pt}
	\label{eq:ES_4}
	{E_{e,t}} = {E_{e,t-1}} + \eta_e^\mathrm{ch}p_{e,t}^\mathrm{ch} - p_{e,t}^\mathrm{dc}/\eta_e^\mathrm{dc}
\end{equation}
\begin{equation}
		\setlength\abovedisplayskip{0pt}
	\setlength\belowdisplayskip{0pt}
	\label{eq:ES_5}
	{E_{e,0}} = E_{e}^\mathrm{init}
\end{equation}
\begin{equation}
	\label{eq:ES_6}
		\setlength\abovedisplayskip{0pt}
	\setlength\belowdisplayskip{0pt}
	{E_{e,\left|\Omega_{\mathrm{T}} \right|}} \ge E_{e}^\mathrm{init}
\end{equation}
 where $p_e^{\max}$ is the power capacity, $[E_e^{\min},E_e^{\max}]$ and $E_{e}^\mathrm{init}$ define the state of charge (SoC) bounds and initial SoC, $\eta_e^\mathrm{ch}$ and $\eta_e^\mathrm{dc}$ are the charge/discharge efficiencies. Variables $p_{e,t}^\mathrm{ch}$, $p_{e,t}^\mathrm{dc}$, and ${E_{e,t}}$ denote the charge/discharge power and the SoC.
 
Frequency regulation load $b\in\Omega_\mathrm{FQR}$ in period $t$ satisfies:
 \begin{equation}
 		\setlength\abovedisplayskip{0pt}
 	\setlength\belowdisplayskip{0pt}
 	\label{eq:FQR_1}
 	p_b^{\min } \le p_{b,t}^\mathrm{cap} \le p_b^{\max }
 \end{equation}
 \begin{equation}
 		\setlength\abovedisplayskip{0pt}
 	\setlength\belowdisplayskip{0pt}
	\label{eq:FQR_2}
	0 \le {p_{b,t}} \le p_{b,t}^\mathrm{cap}
\end{equation}
where $p_b^{\min }$ and $p_b^{\max }$ are the lower/upper bounds of response, and variable $p_{b,t}^\mathrm{cap}$ represents the available frequency regulation capacity that the frequency regulation power ${p_{b,t}} \in [0, p_{b,t}^\mathrm{cap}]$. 

Curtailable load $c\in\Omega_\mathrm{CL}$ in period $t \in \Omega_\mathrm T$ satisfies:   
\begin{equation}
		\setlength\abovedisplayskip{0pt}
	\setlength\belowdisplayskip{0pt}
	\label{eq:CL_1}
	p_c^{\min } \le p_{c,t}^\mathrm{cap} \le p_c^{\max }
\end{equation}
\begin{equation}
		\setlength\abovedisplayskip{0pt}
	\setlength\belowdisplayskip{0pt}
	\label{eq:CL_2}
	0 \le {p_{c,t}} \le p_{c,t}^\mathrm{cap}
\end{equation}
where $p_c^{\min }$ and $ p_c^{\max }$ are the lower/upper bounds of curtailable capacity, while variables $p_{c,t}^\mathrm{cap}$ and ${p_{c,t}}$ represent the curtailable capacity and curtailed power.


During each period, the power balance must be maintained:
\begin{equation}
	\label{eq:PowerBalance_1}
		\setlength\abovedisplayskip{0pt}
	\setlength\belowdisplayskip{0pt}
	\begin{array}{c}
		\sum\nolimits_{g \in {\Omega _\mathrm{SG}}} {{p_{g,t}}}  + \sum\nolimits_{r \in {\Omega _\mathrm{RG}}} {{p_{r,t}}}  + \sum\nolimits_{e \in {\Omega _\mathrm{ES}}} {[ p_{e,t}^\mathrm{dc} -  p_{e,t}^\mathrm{ch}]} 	
		\\ \qquad +\sum\nolimits_{c \in {\Omega _\mathrm{CL}}} {p_{c,t}} + s_{t}^\mathrm{P+} - s_{t}^\mathrm{P-}	
		= \sum\nolimits_{d \in {\Omega _\mathrm{LD}}} {p_{d,t}} 
	\end{array}
\end{equation}
\begin{equation}
		\setlength\abovedisplayskip{-1pt}
	\setlength\belowdisplayskip{0pt}
	\label{eq:PowerBalance_2}
s_{t}^\mathrm{P+}, s_{t}^\mathrm{P-} \ge 0
\end{equation}
where $s_{t}^\mathrm{P+}$ and $s_{t}^\mathrm{P-}$ are the power balance slack variables, and $p_{d,t}$ denotes the load demand $d\in\Omega_\mathrm{LD}$.

To handle the instantaneous power fluctuations $\Delta P$, the maximum frequency deviation criterion must be satisfied:
\begin{equation}
	\setlength\abovedisplayskip{0pt}
	\setlength\belowdisplayskip{0pt}
	\label{eq:frequency_maximum}
	\frac{(\Delta P - \sum\nolimits_{b\in\Omega_\mathrm{FQR}} { {p_{b,t}}} )}{B^{\mathrm{base}}} \le  (1 - \frac{f_{\min }}{f_0})g(H,R,{F_H},T_R) 
\end{equation}
where $\Delta P$ can be set by operation experience, e.g. the percentage of maximum system load \cite{li2021frequency}, and function $g(H,R,{F_H},T_R)$ is defined as:
\begin{equation}
	\setlength\abovedisplayskip{0pt}
	\setlength\belowdisplayskip{0pt}
	\label{eq:g_HRF}
	g(H,R,{F_H},T_R) = \frac{D+1/R}{1+\sqrt{T_{R}^{2} \omega_{n}^{2}-2 \zeta \omega_{n} T_{R}+1} e^{-\zeta \omega_{n} t_{\mathrm{nadir}}}}
\end{equation}
where ${B^{\mathrm{base}}}$, $D$, $f_0$, and $f_{\min }$ are the capacity baseline, damping coefficient, nominal frequency, and minimum frequency. $H$, $R$, $F_H$ and $T_R$ represent system’s inertia time, droop coefficient, fraction of power from high-pressure turbines, and frequency modulation delay coefficient \cite{shi2018analytical}, 
satisfying:
\begin{equation}
	\setlength\abovedisplayskip{0pt}
	\setlength\belowdisplayskip{0pt}
	\label{eq:H_sys}
	{H} = \sum\nolimits_{g \in {\Omega _\mathrm{SG}}} H_g \cdot \frac{{p_g^{\max }x_g}}{B^{\mathrm{base}}}
\end{equation}
\begin{equation}
	\setlength\abovedisplayskip{0pt}
	\setlength\belowdisplayskip{0pt}
	\label{eq:Droop_sys}
	1/{R} = \sum\nolimits_{g \in {\Omega _\mathrm{SG}}} 1/R_g \cdot \frac{{p_g^{\max }x_g}}{B^{\mathrm{base}}}
\end{equation}
\begin{equation}
	\setlength\abovedisplayskip{0pt}
	\setlength\belowdisplayskip{0pt}
	\label{eq:FH_sys}
	{F_H} = \sum\nolimits_{g \in {\Omega _\mathrm{SG}}} {R}/R_g \cdot F_{H,g}x_g
\end{equation}
\begin{equation}
	\setlength\abovedisplayskip{0pt}
	\setlength\belowdisplayskip{0pt}
	\label{eq:TR_sys}
	{T_R} = \sum\nolimits_{g \in {\Omega _\mathrm{SG}}} {R}/R_g \cdot T_{R,g}x_g
\end{equation}
where $\omega_{n}$, $\zeta$, and $t_{\mathrm{nadir}}$ are functions of $H$, $R$, ${F_H}$, and $T_R$. Their specific expressions can be found in \cite{zhang2020modeling}. 

\subsection{Objective Function}
For each period $t\in \Omega_\mathrm T$, the objective function comprises six parts: cost of synchronous generator in day-ahead stage $L_t^\mathrm{SA}$ \eqref{eq:L_SA} and intra-day stage $L_t^\mathrm{SI}$ \eqref{eq:L_SI}, cost of energy storage $L_t^\mathrm{ES}$ \eqref{eq:L_ES}, cost of frequency regulation load $L_t^\mathrm{FQR}$ \eqref{eq:L_FQR}, cost of curtailable load $L_t^\mathrm{CL}$ \eqref{eq:L_CL} and punish cost $L_t^\mathrm{PUN}$ \eqref{eq:L_PUN}.
\begin{equation} 
	\setlength\abovedisplayskip{0pt}
	\setlength\belowdisplayskip{0pt}
	\label{eq:L_SA}
	L_t^\mathrm{SA} = \sum\nolimits_{g \in \Omega_\mathrm{SG}}\left(c_{g}^\mathrm{U} y_{g,t}+c_{g}^\mathrm{D} z_{g,t}\right)
\end{equation}
\begin{equation}
	\setlength\abovedisplayskip{0pt}
	\setlength\belowdisplayskip{0pt}
	\label{eq:L_SI}
	L_t^\mathrm{SI} = \sum\nolimits_{g \in \Omega_\mathrm{SG}}{c_{g}^\mathrm{SG}} p_{g,t}
\end{equation}
\begin{equation}
	\setlength\abovedisplayskip{0pt}
	\setlength\belowdisplayskip{0pt}
	\label{eq:L_ES}
	L_t^\mathrm{ES} = \sum\nolimits_{e \in \Omega_\mathrm{ES}} c_e^{{\rm{ES}}}[p_{e,t}^{{\rm{ch}}} + p_{e,t}^{{\rm{dc}}}]
\end{equation}
\begin{equation}
	\setlength\abovedisplayskip{0pt}
	\setlength\belowdisplayskip{0pt}
	\label{eq:L_FQR}
	L_t^\mathrm{FQR} = \sum\nolimits_{b \in \Omega_\mathrm{FQR}}c_b^{\mathrm{FQR}} {p_{b,t}}
\end{equation} 
\begin{equation}
	\setlength\abovedisplayskip{0pt}
	\setlength\belowdisplayskip{0pt}
	\label{eq:L_CL}
	L_t^\mathrm{CL} = \sum\nolimits_{c \in \Omega_\mathrm{CL}}c_c^{\mathrm{CL}} {p_{c,t}}
\end{equation} 
\begin{equation}
	\setlength\abovedisplayskip{0pt}
	\setlength\belowdisplayskip{0pt}
	\label{eq:L_PUN}
	L_t^\mathrm{PUN} = {c^p}\left( s_{t}^\mathrm{P+}+ s_{t}^\mathrm{P-} \right)
\end{equation} 
where $c_{g}^\mathrm{U}$, $c_{g}^\mathrm{D}$, $c_{g}^\mathrm{SG}$, $c_{e}^{{\rm{ES}}}$, $c_{b}^{{\rm{FQR}}}$, $c_{c}^{{\rm{CL}}}$ are the respective prices for devices, while ${c^p}$ denotes the slack variables' punish price.  

\section{Transform of Maximum Frequency Deviation}
The convex relaxation technique is applied to transform the non-convex maximum frequency deviation constraint \eqref{eq:frequency_maximum}. Initially, the binary variable $x_{g,t}$ in $g(H,R,{F_H},T_R)$
is relaxed to $x_{g,t}\in [0,1]$, and $g(H,R,{F_H},T_R)$ is reformulated as a continuous function $g(\alpha^\mathrm{HG},\alpha _{}^\mathrm{RG},\alpha^\mathrm{FH},\alpha^\mathrm{TR})$, with:
\begin{equation}
	\setlength\abovedisplayskip{0pt}
	\setlength\belowdisplayskip{0pt}
	\label{eq:relationship}
	\alpha _{}^\mathrm{HG} = {H}, \alpha _{}^\mathrm{RG}=1/{R}, \alpha^\mathrm{FH}= F_H/{R}, \alpha^\mathrm{TR}={T_R}/{R}.
\end{equation}

We denote the feasible region of $g(\alpha _{}^\mathrm{HG},\alpha _{}^\mathrm{RG},\alpha^\mathrm{FH},\alpha^\mathrm{TR})$ as set ${\Omega_\mathrm{PA}}$, which is a compact set and can be calculated by the installed synchronous generators' parameters.

The model \eqref{eq:optim_OA} is constructed to assist convex relaxation:
\begin{subequations}
	\label{eq:optim_OA}
	\setlength\abovedisplayskip{0pt}
	\setlength\belowdisplayskip{0pt}
	\begin{gather}
		\label{eq:optim_OA_obj}
		\max\nolimits_{(\alpha _h^\mathrm{HG},\alpha _h^\mathrm{RG},\alpha _h^\mathrm{FH},\alpha _h^\mathrm{TR})\in {\Omega_\mathrm{PA}}} \quad \beta  
		\\	
		\label{eq:optim_OA_cons}
		\text{ s.t.  }
		\beta  \le {L_h}, h = 1,...,{N_\mathrm{HP}}	
	\end{gather}
\end{subequations}
where $(\alpha _h^\mathrm{HG},\alpha _h^\mathrm{RG},\alpha _h^\mathrm{FH},\alpha _h^\mathrm{TR})$ are the selected points within ${\Omega_\mathrm{PA}}$, ${N_\mathrm{HP}}$ is the number of hyperplanes, and
\begin{equation}
	\label{eq:expression_Lh}
	\setlength\abovedisplayskip{0pt}
	\setlength\belowdisplayskip{0pt}
	\begin{array}{l}
		{L_h} =  g(\alpha _h^\mathrm{HG},\alpha _h^\mathrm{RG},\alpha _h^\mathrm{TR},\alpha _h^\mathrm{FH}) 
		\\ \quad+ {\left. {\frac{{\partial g}}{{\partial {\alpha ^\mathrm{HG}}}}} \right|_{\alpha  = {\alpha _h}}}({\alpha ^\mathrm{HG}} - \alpha _h^\mathrm{HG}) +{\left. {\frac{{\partial g}}{{\partial {\alpha ^\mathrm{RG}}}}} \right|_{\alpha  = {\alpha _h}}}({\alpha ^\mathrm{RG}} - \alpha _h^\mathrm{RG}) 
		\\ \quad+ {\left. {\frac{{\partial g}}{{\partial {\alpha ^\mathrm{FH}}}}} \right|_{\alpha  = {\alpha _h}}}({\alpha ^\mathrm{FH}} - \alpha _h^\mathrm{FH}) + {\left. {\frac{{\partial g}}{{\partial {\alpha ^\mathrm{TR}}}}} \right|_{\alpha  = {\alpha _h}}}({\alpha ^\mathrm{TR}} - \alpha _h^\mathrm{TR}).
	\end{array}
\end{equation}
Note that binary variable $x_{g,t}$
has been relaxed to $x_{g,t}\in [0,1]$, it ensures the existence of the partial derivatives in \eqref{eq:expression_Lh}.

The transform method of the maximum frequency deviation constraint \eqref{eq:frequency_maximum} is shown in Algorithm \ref{alg:EA}. The output set, ${\Omega_\mathrm{HP}}$, contains hyperplanes ${L_h}$ ($h = 1,...,{N_\mathrm{HP}}$) that can be the convex relaxation to the non-convex function $g(\alpha^\mathrm{HG},\alpha _{}^\mathrm{RG},\alpha^\mathrm{FH},\alpha^\mathrm{TR})$. Then, equation \eqref{eq:frequency_maximum} can be replaced by constraints with ${L_h}$, see equation \eqref{eq:robust_dispatch_cons_2} for details.
 \begin{algorithm}[htb] 
 	\caption{Transform of maximum frequency deviation.} 
 	\label{alg:EA} 
 	\textbf{Step 1:} 
 	Set initial point $(\alpha _{1}^{HG},\alpha _{1}^{RG},\alpha_1^{FH},\alpha_1^{TR})$ based on \eqref{eq:relationship}, and calculate ${L_1}$ using \eqref{eq:expression_Lh}; initialize ${\Omega_\mathrm{HP}} = \left\{{L_1}\right\}$, ${N_\mathrm{HP}} = 1$; set converge threshold ${\delta _\mathrm{CR}}$. 
 	\\
 	\textbf{Step 2:} 
 	Solve  \eqref{eq:optim_OA}, obtain and denote the optimal solution and value as $(\alpha _{N_{{\rm{HP}}}+1}^{HG},\alpha _{N_{{\rm{HP}}+1}}^{RG},\alpha _{N_{{\rm{HP}}+1}}^{TR},\alpha _{N_{{\rm{HP}}}+1}^{FH})$ and $\hat{\beta}_{N_\mathrm{HP}}$. \\	
 	\textbf{Step 3:} If $\hat{\beta}_{N_\mathrm{HP}} - {\max _{h \in [1,{N_{{\rm{HP}}}}+1]}}\left[ {g(\alpha _h^{HG},\alpha _h^{RG},\alpha _h^{FH},\alpha _h^{TR})} \right]$ $ \ge {\delta _\mathrm{CR}}$, update ${N_\mathrm{HP}} = {N_\mathrm{HP}} + 1$ and ${\Omega_\mathrm{HP}} = {\Omega_\mathrm{HP}}\cup \left\{{L_{N_{{\rm{HP}}}}}\right\}$, and return to Step 2;
 	else, go to Step 4.\\
 	\textbf{Step 4:} End iterations, and output set ${\Omega_\mathrm{HP}}$. 
 \end{algorithm}
 
\section{Day-ahead robust dispatch framework}
\subsection{Framework Composition}\label{Framework Composition}
\begin{figure}[htb]
	\centering
	\includegraphics[width=7.5cm]{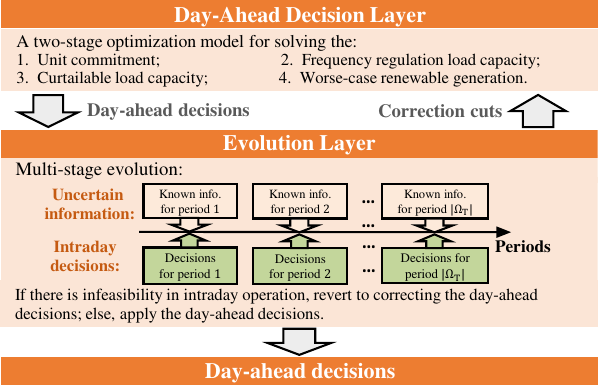}
	\caption{The evolution-based robust dispatch framework.}
	\label{Fig:decision_framework}
\end{figure}
The evolution-based robust dispatch framework consists of the day-ahead decision layer and the evolution layer. Fig. \ref{Fig:decision_framework} illustrates the relationship between these two layers.

\subsection{Day-Ahead Decision Layer} 
This layer is responsible for determining decisions for unit commitment (UC) ${x_{g,t}}$, frequency regulation load capacity $p_{b,t}^\mathrm{cap}$, and curtailable load capacity $p_{c,t}^\mathrm{cap}$.
The $min$-$max$-$min$ structured robust dispatch model \eqref{eq:robust_dispatch} is constructed, which is a two-stage problem solvable efficiently by iterative algorithms such as column-and-constraint generation (C\&CG).
\begin{subequations}
	\label{eq:robust_dispatch}
	\setlength\abovedisplayskip{0pt}
	\setlength\belowdisplayskip{0pt}
	\begin{gather}
		\label{eq:robust_dispatch_obj}
		\begin{aligned}
			\min\nolimits_{x_{g,t},y_{g,t},z_{g,t},\atop p_{b,t}^\mathrm{cap},p_{c,t}^\mathrm{cap}}& \sum\nolimits_{t \in {\Omega _\mathrm{T}}}(L_t^\mathrm{SA} + L_t^\mathrm{FQR} + L_t^\mathrm{CL}) \\ + \max\nolimits_{{p_{r,t}}} &\min{\sum\nolimits_{t \in {\Omega _\mathrm{T}}} } \left( L_t^\mathrm{SI}+L_t^\mathrm{ES}+L_t^\mathrm{PUN} \right)
		\end{aligned}
		\\	
		\notag
		\text{ s.t. for $\forall t , g, r, E, b,  c,  l $: }
		\\
		\label{eq:robust_dispatch_cons_1}
		\eqref{eq:SG_1}-\eqref{eq:PowerBalance_2}, \, \eqref{eq:H_sys}-\eqref{eq:TR_sys}
		\\
		\label{eq:robust_dispatch_cons_2}
		\frac{(\Delta P - \sum\nolimits_{b\in\Omega_\mathrm{FQR}} { {p_{b,t}}} )}{B^{\mathrm{base}}} \le (1 - \frac{f_{\min }}{f_0}){L_h},
		{L_h} \in {\Omega_\mathrm{HP}}
		\\
		\label{eq:robust_dispatch_cons_3}
		\{{x}_{g,t},{p}_{b,t}^\mathrm{cap}, {p}_{c,t}^\mathrm{cap} \}\in {\Omega_\mathrm{CUT}}
	\end{gather}
\end{subequations}
where constraints \eqref{eq:SG_1}-\eqref{eq:SG_3}, \eqref{eq:FQR_1}, \eqref{eq:CL_1}, and \eqref{eq:robust_dispatch_cons_3}  serve as the first-stage constraints, and the others are second-stage constraints. ${\Omega_\mathrm{CUT}}$ is the correction set, and can be intialized as ${\Omega_\mathrm{CUT}}=\left\{\eqref{eq:SG_1}-\eqref{eq:SG_3},\eqref{eq:FQR_1},\eqref{eq:CL_1}\right\}$.

Solving this model yields optimal day-ahead decisions ${\hat{x}_{g,t}}$, ${\hat{p}^\mathrm{cap}_{b,t}}$, ${\hat{p}^\mathrm{cap}_{c,t}}$ ($\forall g,b,c,t $). We can also obtain adverse scenarios for renewable generation ${p_{r,t}}$,  denoted by $\tilde{p} _{r,i} = \{\tilde{p}_{r,t,i}\}_{t=1}^{\left|\Omega_{\mathrm{T}} \right|}$ for the $i$-th adverse scenario, with each $\tilde{p}_{r,t,i}$ being the adverse scenario for renewable generation $r$ in period $t$. The set of adverse scenarios is denoted ${\Omega_\mathrm{W}}$, with quantity $\left|\Omega_{\mathrm{W}}\right|$.

\subsection{Evolution Layer}
The two-stage model \eqref{eq:robust_dispatch} assumes that renewable generation across all periods is observed simultaneously, which is not practical for real-time operation. Under adverse scenarios, this can result in infeasibility for intraday decisions \cite{kang2023day}.

To address this, we develop a multi-stage optimization problem to simulate the intraday operation under adverse scenario set $\Omega_\mathrm W$, a process we term ``evolution". Assessing the feasibility of intraday decisions is valuable for analyzing and making necessary adjustments for day-ahead decisions. 

For each adverse scenario $\tilde{p} _{r,i} \in \Omega_\mathrm W$ and period $t \in \Omega_\mathrm T$, we formulate the following optimization model:
\begin{subequations}
	\label{eq:CE}
	\setlength\abovedisplayskip{0pt}
	\setlength\belowdisplayskip{0pt}
	\begin{gather}
		\label{eq:CE_obj}
		\begin{aligned}
			\min (L_t^\mathrm{SI}+L_t^\mathrm{ES}+L_t^\mathrm{PUN}) 
		\end{aligned}
		\\
		\label{eq:CE_cons_1}
		\text{ s.t.  } \eqref{eq:SG_4}-\eqref{eq:SG_6}, \eqref{eq:CL_2}: \alpha_{g,t}^{\pm},\beta_{g,t}^{+},\beta_{g,t}^{-},\gamma_{c,t}^{\pm}
		\\
		\label{eq:CE_cons_2}
		\eqref{eq:ES_1}-\eqref{eq:ES_6}, \eqref{eq:FQR_2},\eqref{eq:PowerBalance_1}-\eqref{eq:PowerBalance_2}
		\\
		\label{eq:CE_cons_2_add}
		\frac{(\Delta P - \sum\nolimits_{b\in\Omega_\mathrm{FQR}} { {p_{b,t}}} )}{B^{\mathrm{base}}} \le (1 - \frac{f_{\min }}{f_0}){L_h},
		{L_h} \in {\Omega_\mathrm{HP}}
		\\
		\label{eq:CE_cons_3}
		{x_{g,t}} = \hat{x}_{g,t},\,{p_{b,t}} = \hat{p}_{b,t}^\mathrm{cap},\, {p_{c,t}} = \hat{p}_{c,t}^\mathrm{cap},\,
		{p_{r,t}} = \tilde{p}_{r,t,i}	
	\end{gather}
\end{subequations}
where dual variables $\alpha_{g,t}^{\pm},\beta_{g,t}^{+},\beta_{g,t}^{-},\gamma_{c,t}^{\pm}$ are associated with the constraints. Let $\hat{L}_{t,i}^\mathrm{PUN}$ denote the optimal value of $L_t^\mathrm{PUN}$ for period $t$ under scenario $\tilde{p}_{r,t,i}$, and let the dual variable values be denoted by $\hat{\alpha}_{g,t}^{\pm},\hat{\beta}_{g,t}^{+},\hat{\beta}_{g,t}^{-},\hat{\gamma}_{c,t}^{\pm}$. 

After solving model \eqref{eq:CE} for $\forall \tilde{p} _{r,i} \in \Omega_\mathrm W$ and $\forall t \in \Omega_\mathrm T$, these evolution results can be used to verify and improve the day-ahead decisions, further detailed in Section \ref{Soultion Algorithm}. 

\section{Solution Algorithm for the Framework}
\label{Soultion Algorithm}

\subsection{Controlled evolution-based day-ahead decision correction}
Noted that $\hat{L}_{t,i}^\mathrm{PUN}>0$ indicates that the current decisions cause operation infeasibility under adverse scenario $\tilde{p}_{r,i}$. For adverse scenarios $\tilde{p} _{r,i} \in \Omega_\mathrm W$ and periods $t \in \Omega_\mathrm T$ where $\hat{L}_{t,i}^\mathrm{PUN}>0$, the cutting planes based on the gradient of Lagrangian function for \eqref{eq:CE} are constructed as follows:
\begin{equation}
	\label{eq:CE_correction}
	\setlength\abovedisplayskip{0pt}
	\setlength\belowdisplayskip{0pt}
	\begin{array}{l}
		\sum\nolimits_{g \in \Omega_\mathrm{SG}} {\left[ \begin{array}{l}
				\hat{\alpha}_{g,t}^{-}p_g^{\max }  + \hat{\alpha}_{g,t}^{+}p_g^{\min } \\
				+  \hat{\beta}_{g,t}^{+}( p_g^{\max }-{RD_g} )\\
				+  \hat{\beta}_{g,t}^{-}(p_g^{\max }-{RU_g})
			\end{array} \right] \cdot \left({x_{g,t}}-{\hat{x}_{g,t}} \right)} \\
		+ \sum\nolimits_{c\in\Omega_\mathrm{CL}} {\hat{\gamma}_{c,t}^{-} \left(   p_{c,t}^\mathrm{cap} - \hat{p}_{c,t}^\mathrm{cap} \right)}  \le  -\hat{L}_{t,i}^\mathrm{PUN}
	\end{array}
\end{equation}

The theoretical derivation and detailed explanation of \eqref{eq:CE_correction} can be found in \cite{cadillack_CE_2025}. By adding \eqref{eq:CE_correction} into ${\Omega_\mathrm{CUT}}$ and re-solving \eqref{eq:robust_dispatch}, the day-ahead decisions can be improved to ensure feasibility.

\subsection{Day-ahead robust dispatch algorithm}
\begin{figure}[htb]
	\centering
	\includegraphics[width=6cm]{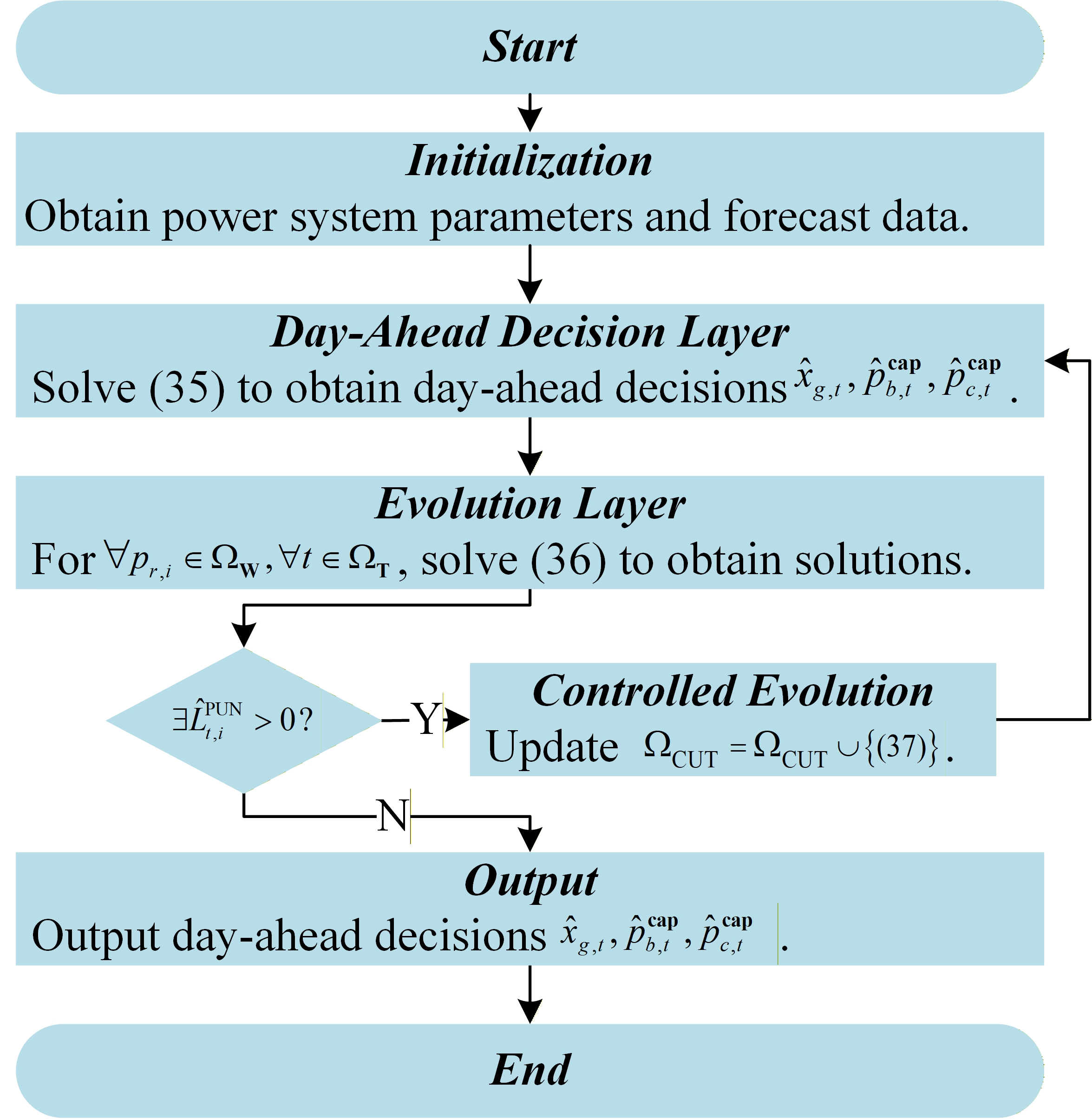}
	\caption{The dispatch algorithm for the framework.}
	\label{Fig:solution_algorithm}
\end{figure}

Corresponding to the proposed framework, the day-ahead robust dispatch algorithm is shown in Fig. \ref{Fig:solution_algorithm}.

\section{Case Studies}
\subsection{Set-Up}
\begin{figure}[htb]
	\centering
	\includegraphics[width=7.5cm]{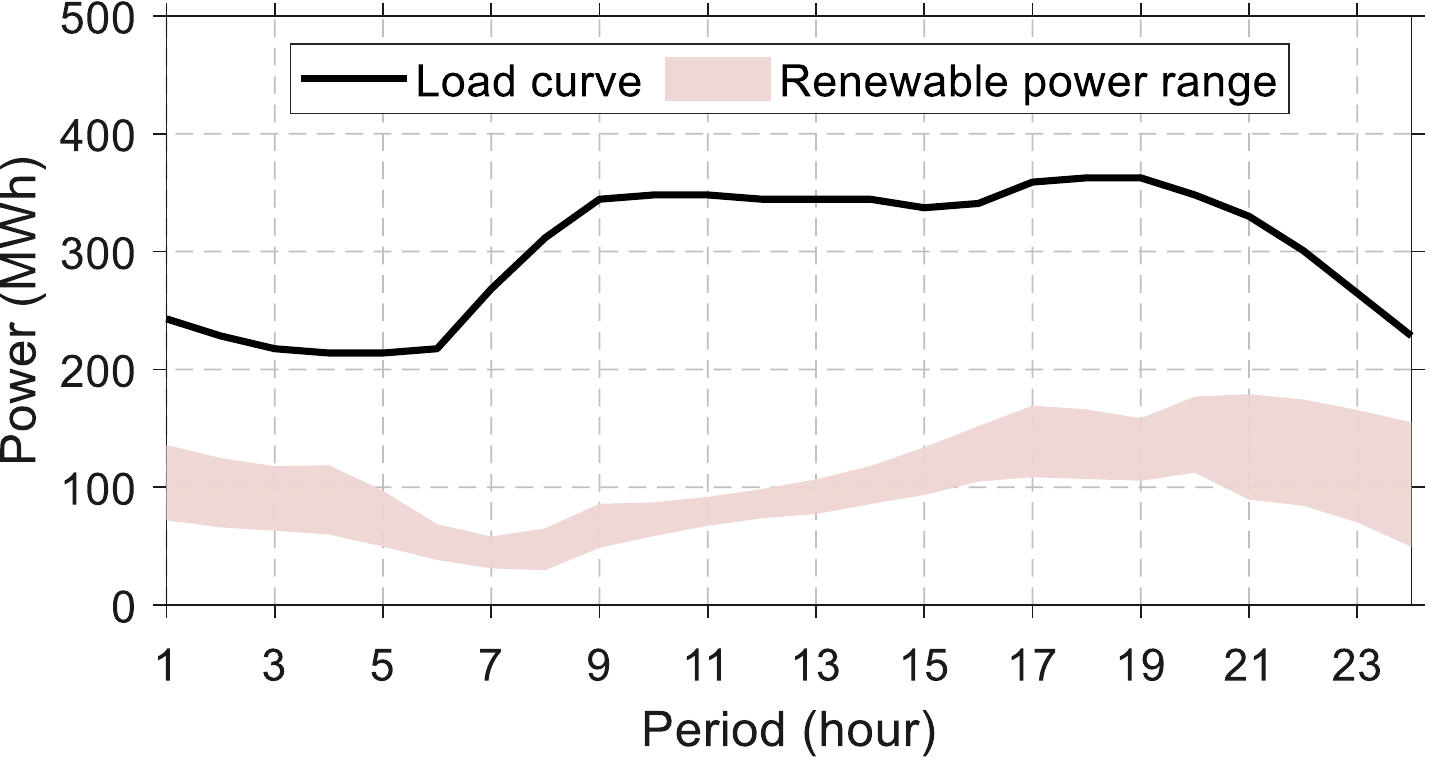}
	\caption{Renewable power range and total load demand curve.}
	\label{Fig:Load_renewable_curve}
\end{figure}

A modified IEEE 14-bus system is used for case studies, including four synchronous generators (denoted as G1-G4), an energy storage station, renewable generation, frequency regulation load, curtailable load, and several load demands. The key parameters of the generators and energy storage are presented in Tables \ref{Tab:Parameters of the generators} and \ref{Tab:Parameters of the energy storage}. Fig. \ref{Fig:Load_renewable_curve} depicts the renewable power range and total load curve.
The upper bounds for frequency regulation capacity $p_b^{\max }$ is set to 30 MW. The upper bound of curtailable load capacity $p_c^{\max }$ and the maximum instantaneous power fluctuations $\Delta P$ are both set to 10\% of the total load. 
Parameters $f_0$,  $f_{\min }$, and $D$ are set to 50Hz, 49.7Hz, and 2MW/Hz. The punish price ${c^p}$ is set to 1000 \$/MW.
%
%
\begin{table}[htb]
	\footnotesize
	\caption{Parameters of the synchronous generators.}
	\label{Tab:Parameters of the generators}
	\setlength{\tabcolsep}{3pt}
	\centering
	\renewcommand\arraystretch{0.5} 
	\begin{tabular}{p{72pt}<{\centering} m{38pt}<{\centering} m{38pt}<{\centering} m{38pt}<{\centering} m{38pt}<{\centering}}
			\toprule
			& G1 &  G2 &	G3 &  G4 \\[0pt]
			\midrule
			$[	p_g^{\min },p_g^{\max }]$ (MW) &[100,200]	&[100,200] &[80,150] & [10,50] \\[5pt]
			${RD_g}$, ${RU_g}$ (MW/h) &75&75&50&40\\[3pt]
			$H_g$ (s) &	7 & 7 &5 &4 		\\[3pt]
			$1/R_g$ (MW/h)&	25&	25&	20 & 20 \\[3pt]
			$F_H$ & 0.3 &	0.3  &	0.2 & 0.2 \\[3pt]
			$T_R $  (s) &	6 &	5 &	4 & 3 		\\[3pt]
			$c_{g}^\mathrm{U}$, $c_{g}^\mathrm{D}$ (\$)&	1500 &	1500 &	1000 & 500		\\[3pt]
			$c_{g}^\mathrm{SG}$ (\$/MW)& 13.29 &	13.29 & 15.47 &	14.50
			\\[0pt]
			\bottomrule
		\end{tabular}
	\end{table}
	\begin{table}[htb]
		\footnotesize
		\caption{Parameters of the energy storage station.}
		\label{Tab:Parameters of the energy storage}
		\setlength{\tabcolsep}{3pt}
		\centering
		\renewcommand\arraystretch{0.6} 
		\begin{tabular}{p{45pt}<{\centering} m{45pt}<{\centering} m{45pt}<{\centering} m{45pt}<{\centering} m{45pt}<{\centering}}
				\toprule
				& $p_e^{\max}$ &  $[E_e^{\min},E_e^{\max}]$ &	$\eta_e^\mathrm{ch},\eta_e^\mathrm{dc}$ &  $E_{e}^\mathrm{init}$ \\[0pt]
				\midrule
				Value & 10MW	&[2,18]MWh &0.95 & 10MWh
				\\[0pt]
				\bottomrule
		\end{tabular}
\end{table}
\begin{figure}[htb]
	\centering
	\includegraphics[width=7.8cm]{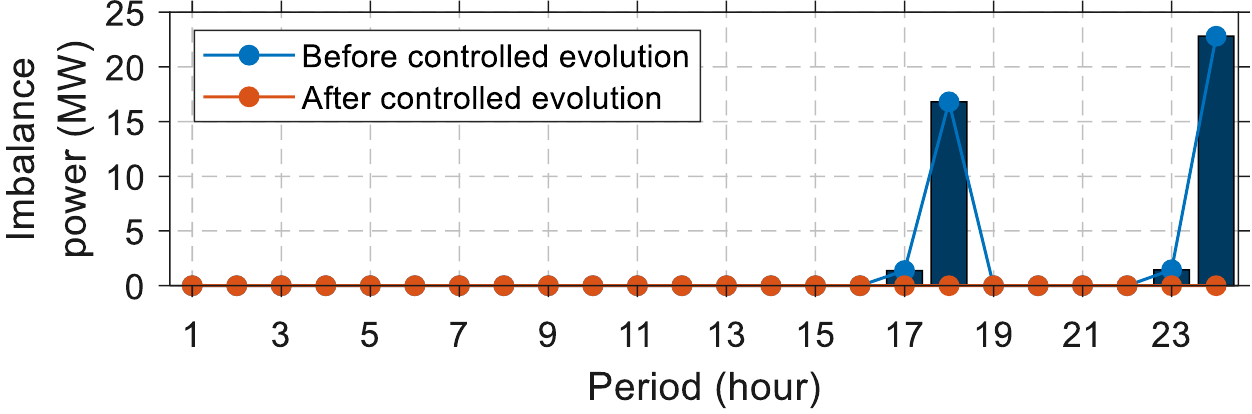}
	\caption{Imbalance power before and after controlled evolution.}
	\label{Fig:Imbalance_power}
\end{figure}
\begin{figure}[htb]
	\centering
	\includegraphics[width=7.8cm]{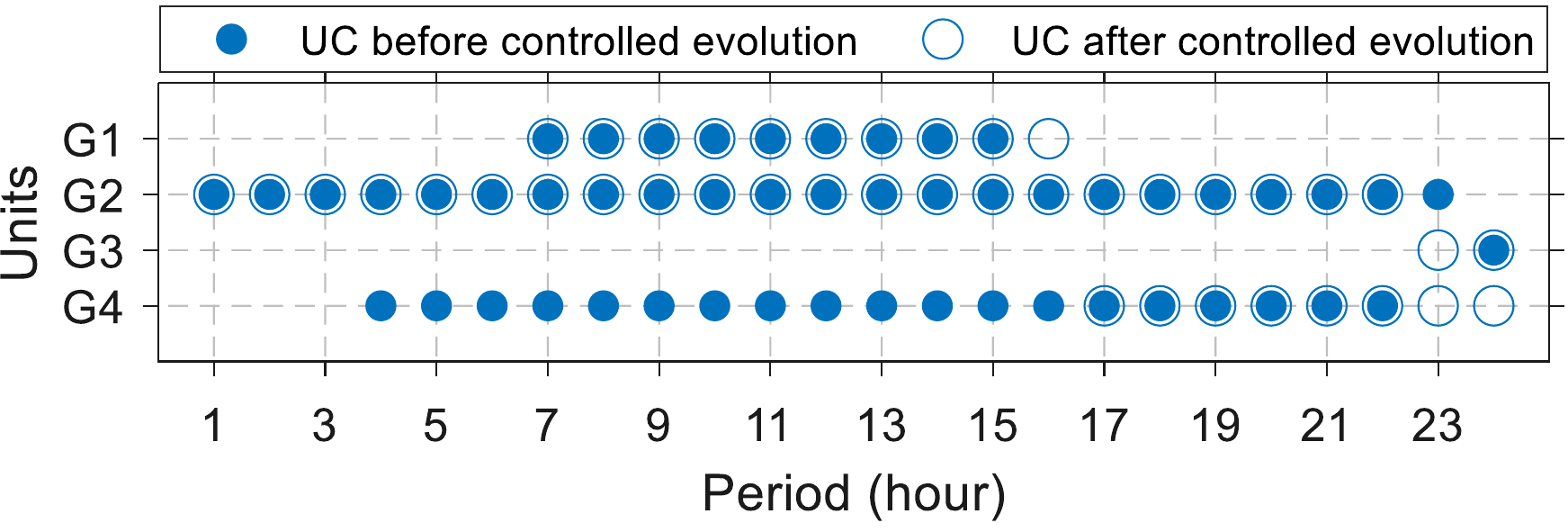}
	\caption{UC before and after the controlled evolution.}
	\label{Fig:UC}
\end{figure}
\begin{figure}[htb]
	\centering
	\includegraphics[width=7.8cm]{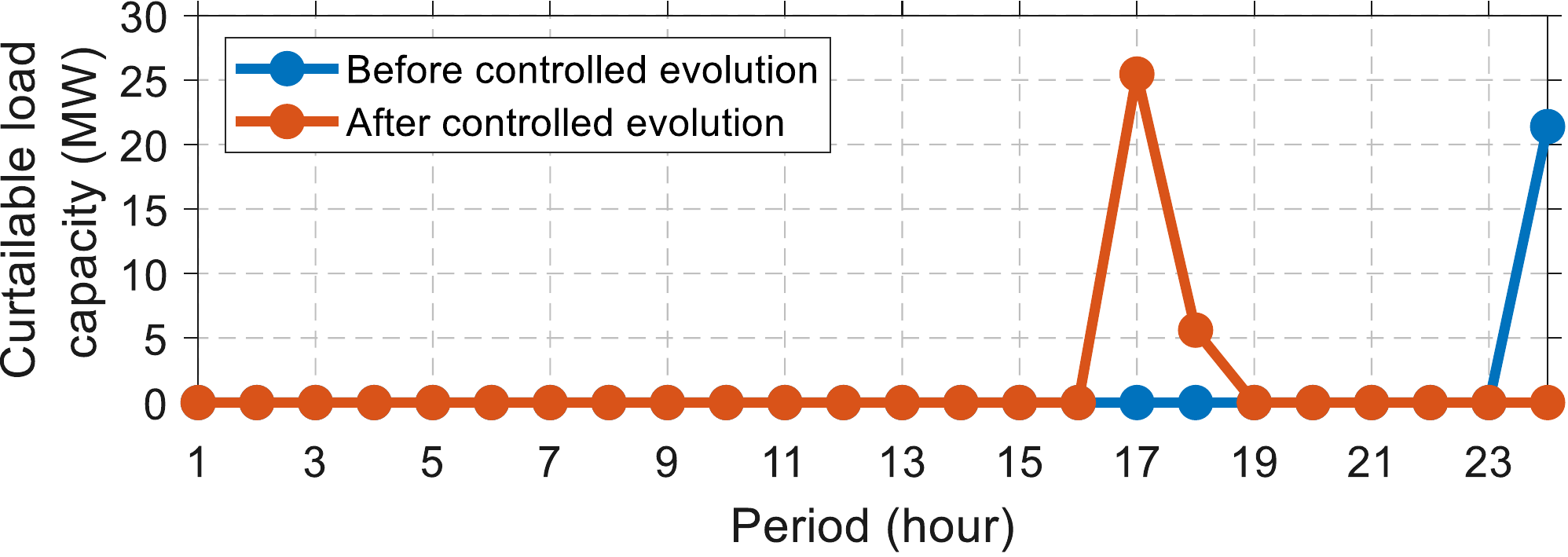}
	\caption{Curtailable load capacity.}
	\label{Fig:Curtailable_capacity}
\end{figure} 
\begin{figure}[htb]
	\centering
	\includegraphics[width=7.8cm]{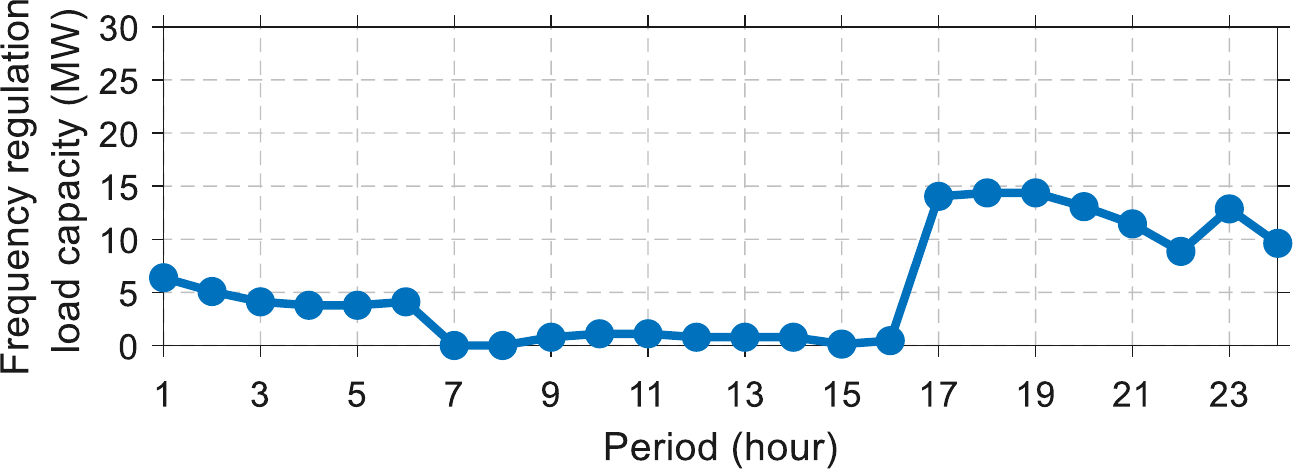}
	\caption{Frequency requlation load capacity.}
	\label{Fig:_Frequency_requlation_load_capacity}
\end{figure} 
\begin{table*}[h]
	\footnotesize
	\centering
	\caption{Maximum frequency deviation after the instantaneous power deviation  $\Delta P$.}
	\label{Tab:Frequency_index}
	\tabcolsep=0.012\linewidth
	\renewcommand\arraystretch{0.6} 
	\begin{tabular*}{\linewidth}{@{}lcccccccccccc@{}}
		\toprule
		Periods &1&2&3&4&5&6&7&8&9&10&11&12	
		\\
		\midrule
		Maximum frequency deviation & 0.2962 &   0.2979 &   0.2992 &   0.2996   & 0.2996 &   0.2992 &   0.2468 &   0.2803 &   0.2975&    0.2972 &   0.2972 &   0.2975\\
		\midrule
		Periods &13&14&15&16 &17&18&19&20&21&22&23&24 \\
		\midrule
		Maximum frequency deviation & 0.2975 &   0.2975 &   0.2979  &  0.2977 &0.2966 &   0.2963 &   0.2963  &  0.2977 &   0.2994  &  0.3021 &   0.3184 &   0.3210 \\
		\bottomrule
	\end{tabular*}
\end{table*}

\subsection{Calculation results}
In the following, the initial day-ahead decisions obtained from the day-ahead decision layer are referred to as "before the controlled evolution", while the corrected day-ahead decisions are labeled as ``after the controlled evolution". 

Fig. \ref{Fig:Imbalance_power} illustrates the average imbalance power of the initial and corrected day-ahead decisions when applied to adverse scenarios $\tilde{p} _{r,i} \in \Omega_\mathrm W$. For the initial decisions, there are power imbalances during periods 17-18 and 23-24, while the corrected decisions avoid the imbalances.

Additionally, Fig. \ref{Fig:UC} and Fig. \ref{Fig:Curtailable_capacity} display the UC and curtailable load capacity before and after controlled evolution. 
After correction, additional generators are started in periods 16 and 23-24 to address imbalance risks, and curtailable load capacity increases in periods 17-18, helping to mitigate imbalances.

For frequency security, Fig. \ref{Fig:_Frequency_requlation_load_capacity} displays frequency regulation load capacity decisions, and Table \ref{Tab:Frequency_index} presents maximum frequency deviations following an instantaneous power deviation $\Delta P$. Most of the maximum frequency deviations stay within the preset range $f_0-f_{\min }$, ensuring system frequency security. For the few periods 22-24 where deviations exceed $f_0-f_{\min}$, presetting a slightly larger power deviation than $\Delta P$ in equation \eqref{eq:frequency_maximum} could improve frequency robustness.

For a discussion on the computational time, and the comparison with affine policy-based dispatch method, please refer to \cite{cadillack_CE_2025}, which also highlight the advantages of our method.

\section{Conclusions}
To address challenges of satisfying maximum frequency deviation in high-renewable power systems and incorporate causality into decision-making, we propose a controlled evolution-based day-ahead robust dispatch method. The convex relaxation technique is introduced to transform the maximum frequency deviation to facilitate optimization. Additionally, we present a evolution-based day-ahead robust dispatch framework that incorporates causality. Furthermore, the controlled-evolution based algorithm is developed to solve this framework efficiently. Results from the modified IEEE 14-bus system show that our method significantly improves the feasibility of day-ahead decisions in intraday operations and effectively enhances frequency security.

\section*{Acknowledgment}
This work is supported by Science and Technology Foundation of State Grid Corporation of China (Grant Number: 5100-202355765A-3-5-YS).


\ifCLASSOPTIONcaptionsoff
\newpage
\fi
\bibliographystyle{IEEEtran}
\bibliography{IEEEabrv,mybib}

\end{document}